# Conditioning of Superconductive Properties in Graph-Shaped Reticles

*M. Lucci[1], D. Cassi[2], V. Merlo[1], R. Russo[3], G. Salina[4], M. Cirillo[1*]*

[1]Dipartimento di Fisica and *MINAS Lab*, Università di Roma *Tor Vergata*, I-00133 Roma, Italy

[2]Dipartimento di Scienze Matematiche, Fisiche ed Informatiche, Università di Parma, I-43124 Parma, Italy

[3]CNR-IMM Via Pietro Castellino, 111, 80131 Napoli (Italy)

[4]Istituto Nazionale di Fisica Nucleare, Sezione Roma "Tor Vergata", Via della Ricerca Scientifica 1, 00133 Roma

## Abstract

We report on phenomena observed in planar integrated networks obtained connecting superconducting island by Josephson tunnel junctions. These networks, identifiable as tree-like graphs, have branches consisting of series arrays of Josephson junctions which can be individually current biased and characterized. Both Josephson supercurrents and gap parameters of the arrays embedded in the graph structures display properties significantly different from those of "reference" arrays fabricated on the same chips and having identical geometrical shape. The temperature and magnetic field dependencies of the Josephson current of the embedded arrays both show a singular behavior when a critical value is reached by the Josephson characteristic energy. The gap parameter of the junctions generating the embedded arrays is higher than that of the junctions forming the reference geometrical arrays.

(*) Corresponding author, *cirillo@roma2.infn.it*

# 1) INTRODUCTION

The possibility that in a double comb-shaped network specific effects could be observable in carriers distribution (Cooper pairs) over the superconductive islands generating the reticles was first reported by Burioni et al [1,2]: these authors, and others [3], predicted a Bose-Einstein Condensation BEC for peculiar graph topologies. More recently, the specific issue of condensation on graphs has been also attacked by several groups from a mathematical and statistical point of view [4-7]. As far as the experimental side is concerned, several papers have reported on results consistent with the theoretical predictions [8,9,10]. In Figure 1a we show a sketch of a graph structure having the shape of a double comb, the structure first analyzed in ref. 1: here the dots are the superconductive islands and the lines represent the connections (through Josephson elements). In Figure 1b a typical experimental realization of such a structure with superconductive pads connected by Josephson junctions is shown. It is worth noting that the volumes of the superconducting islands are engineered to contain the same amount of Cooper pairs available for tunnelling in each junctions. Thus, an island on which four junctions are present has twice the volume of an island originating only two junctions.

The predictions of the theory are concerned with thermal hopping of bosons (Cooper pairs), via Josephson tunnelling, between superconductive islands which can take place when the thermal excitations energy is comparable with Josephson potentials all over the graph array distribution [1,8]. Considered that experiments on samples fabricated in conventional superconductive technologies are typically performed at about *4.2 K* and below, this condition imposes limitations on the amplitude of the Josephson currents $I_c$ and relative zero-bias energy $E_J = \Phi_0 I_c / 2\pi$ [7,9], where $\Phi_0 = 2.07 \times 10^{-15}$ Wb is the flux quantum. Therefore the estimate of the currents, for the zero-bias case, is given by $I_c = (2\pi/\Phi_0) k_B T$. Substituting numerical values, with $k_B = 1.38 \times 10^{-23}$ J/K (Boltzmann's constant) and *T=4.2 K,* in the last equation we get *$I_c$=176 nA*. We conclude that a current of few hundred nanoampères all over the junctions of the arrays is necessary in order to be in the conditions of the theoretical model.

The predictions of the theory were probed by measuring the current-voltage characteristics of the arrays of Josephson junctions representing branches of the networks where the specific topological effects in carrier distribution were expected [8, 9, 10]. This "probing" technique does allow setting the energy of the current-biased array-branches to the same order of magnitude of the thermal excitations, because the external bias current feeding the junctions, generates a tilt of the potential [11] which lowers the Josephson energy barrier according to the equation [10] $\Delta E = 2E_j(\sqrt{1-\rho^2} - \rho \cos^{-1}\rho)$ where $\rho = I/I_c$ is the bias current fed through the junction *(I)* normalized to its maximum Josephson current *($I_c$)*. It is straightforward to see that, close to the maximum of the Josephson current, namely when $\rho \to 1$ *($I \to I_c$)*, this energy can become comparable with the thermal excitations at *4.2K*, no matter what the value of the maximum current is. Still, probing one array at time, in the current unbiased arrays branches of the graphs (e.g. the lateral fingers in the sketch of Fig. 1a) the thermal energy remains much lower than the Josephson coupling energy. In these conditions the requirement that the "hopping" thermal energy of the carriers should be of the order of the energy of the potential separating the superconductive islands does not hold all over the junctions of the arrays. In spite of this limitation noticeable effects have been observed in measurements [8, 9, 10] and indeed none of the recorded features has been found in contraddiction with the theoretical predictions.

In order to prevent the limitation described in the past paragraph we have performed experiments in which an external magnetic field, or the temperature, can lower the Josephson currents (and consequently energies) of all the junctions of the graph array structures. In what follows we report on the results of such experiments which have enabled us to observe effect much more evident than those reported before. Morever, other results are observed which go beyond the specific predictions of the theoretical model. It is found, in particular, that not only the Josephson current, but even the gap energy of the junctions embedded in the graph structures are different from those of the junctions generating reference arrays.

## 2) RESULTS

The samples tested in the experiments were designed following the procedures employed at *Seeqc* (Elmsford, NY,USA) where the chips were fabricated in a niobium trilayers technology for a *100 A/cm$^2$* current density process [12]. Typical result of the fabrication procedure is shown in Fig. 1b where one can see a portion of a double comb array: in particular, the backbone line is visible, along the horizontal, central direction of the aligned crosses.

The areas of the juctions are squares having *3μm* side and are clearly visible in the photo. A specific difference between the present design and previous ones [8-10] is the fact that we have "isolated" the arrays from the large contact pads by using normal thin film contacts at the ends of the arrays for feeding current through them and reading voltages. This specific fabrication step was also followed in order to exclude perturbing effects at the ends of the arrays generated by the large superconducting contact pads. Most of the measurements were performed at *4.2K* keeping the samples in liquid helium and temperature dependencies were performed in helium vapours. Cryoperm shielding was used to protect the samples from spurious magnetic fields and electromagnetic noise, while external magnetic fields were applied in the plane of the barrier of the junctions (the direction is indicated in Fig. 1b) by solenoids surronding the samples . The data were acquired in a system based on LabView software and statistical/fitting data analysis was worked out by MATLAB and other scientific packages. The results herein reported are very representative of those that we have obtained on *16* samples.

We will herein focus on the results obtained on "double comb" graph reticles. In these arrays we have mainly characterized two "branches" which are those indicated in Fig. 1a by the contact pads at the ends: the backbone array (horizontal in figure), and the central "finger" array (vertical in the figure). The latter is a series array represented by two fingers aligned on the two sides of the "backbone" line and it is indeed a "double" finger, but we will refer to it herein just as "central finger" array. Along the backbone array all the superconductive islands have coordination number equal to 4 (each island is connected to 4 neighbours via Josephson junctions, follow Fig. 1a,b). As shown in Fig. 1a current is fed from the ends of the arrays where voltage is measured in a four probe configuration. The "finger" arrays (double fingers arrays indeed) have only one island (the one in common with the backbone) having coordination number equal to 4 while all the other islands have coordination number equal to 2.

The backbone arrays embedded in the graph contains *200* junctions (*4* for each island), but biasing the arrays as shown in Fig. 1a we only feed current through *100* junctions because those connecting the backbone islands to the fingers are not biased. While the theory was worked out in the thermodynamical limit we must specify that we do have boundaries. In particular, the final islands of the backbone arrays have coordination number *3* while the final units of the fingers have coordination number *1*. As visible in Fig. 1b along the back bone lines we alternate islands having a cross shape with others having a square shape. The volume of the islands, however, "normalized" to the number of junctions present on these is the same for all the islands: an island having four junctions on it has twice the volume of an island having two junctions. In Fig. 2b it is evident that the crosses-shaped islands of the backbone have smaller planar dimensions than the square-shaped ones, however, their thickness is larger and so the volumes are the same. In all figures herein presented we have a label indicating the specific sample on which the measurement was obtained: this has been necessary due to the vast amount of data that we have collected.

Along with the two just mentioned arrays we characterize their "reference" arrays. These reference arrays have the same geometrical shape of the ones embedded in the graphs that we test, but have a different topological structure. The backbone reference array has all the islands with coordination number equal to two because all the islands are missing the connections to the fingers, whereas in the double finger reference arrays the central island has coordination number equal to 2, because it is not connected to the rest of the backbone array. These differences enable us to distinguish geometrical effects from topological ones between graph-embedded and isolated arrays because the current biasing conditions and the geometry are the same in the two types of arrays that we compare.

In Fig. 2a we show, on the same horizontal and vertical scales, the current-voltage characteristics of two arrays: the backbone array embedded in the graph structure (black curve) and its reference array (red curve). These arrays both contain *100* junctions in series and two things are evident when comparing their characteristics: the Josephson currents of the graph-embedded backbone array are higher, over the whole voltage span of those of the reference array. An average over the whole voltage span reveals that the backbone array has a current higher of *1.4 $\mu A$* with respect to the reference array when averaged over the whole voltage span. This corresponds roughly to *10%* of the average current of the reference array. Phenomenona similar to those visible in Fig. 2a have been reported previously [8,9,10], but the fact that we now note is that even the gap voltages of the junctions of the array embedded in the graph structure result higher than those of the reference array. In Fig. 2b we have a zoom of the data in Fig. 2a showing that every junction of the embedded backbone array has a gap voltage higher than that of the reference array: in this specific case the increase is *75$\mu V$* for each junction. As we see in Fig. 2b the individual contributions of each junction sum because of the series connection and, for all the series-connected junctions of the array we reach a value of about *7.5 mV* which makes the substantial difference visible at the gap sum in Fig. 2a. This gap increase corresponds to 3% of the gap of the junctions of the reference array and therefore is not directly, quantitatively, linkable to Josephson current increase, however this phenomenon, it is also strictly related to the topology. All the reference arrays that we tested had the same voltage and therefore gap increase can only be attributed to the specific topological configuration, as we also checked on other graph structures [13]. In all the arrays we tested the increase/per junction of the gap in the backbone arrays ranged in the *(40-80) $\mu V$* range.

Note that in the very first experiments performed on comb arrays (Ref.8) the quality of the arrays was definitely not very good. However, the noticeable effect on the gaps (increase for the backbone and finger arrays) was clearly visible at *50 mK*. The chips of Ref. 9 contained an error in the design: the backbone "topology" was not uniform since it was a sequence of *2* and *4* coordination number. A peculiar "hybrid" fabrication technique was used for the samples of Ref. 10 . We judged that for the present experiments a very "standard" Nb trilayer technology would be advisable. In any case, there are several improvements in the present design with respect to the previous ones. The most relevant one, is the already mentioned use normal of contacts at the ends of the arrays for reading voltages and current-biasing. However, we also put some specific extra contacts probes (made of normal thin films) for testing specific junctions located on the finger arrays. The Josephson junctions junctions are planar *3μm*-side square junctions and their quality is very acceptable with with a $V_m$ *(@2mV) = 70 mV* and a gap sum equal to *2.7 mV*.

As we said above a significant difference in the gap sum like that shown in Fig. 2a was also visible in a previous paper [8], however, attention was not dedicated to clarify this specific phenomenon whose origin was attributed, although not declared, to failures in the fabrication process: due to these failures the reference arrays could have some shorted junctions and therefore have a lower gap-sum. We have no doubts now that both the arrays (graph-embedded and reference) have the same number of junctions because we have counted their number, one by one, from the gap jumps in the current-voltage characteristics. The real effect is that each junction of the graph-embedded backbone array has a slightly larger gap, as shown in Fig. 2b. Note that the subgap current was identical for embedded and reference arrays (as well as the normal state resistances) ,as shown in Fig. 2b, and surely do not justify the different Josephson currents of the two arrays. The noise fluctuations, measured on the top of the Josephson currents were of the order of *100 nA* while on the subgap resistances the value was a factor *10* below this value.

Let us characterize now the observed differences in the Josephson currents of the two arrays (embedded backbone and its reference) as a function of applied magnetic field. As mentioned in the introduction this test allows decreasing the Josephson energy all over the arrays so that it becomes comparable to the thermal excitations in all the junctions, biased and unbiased. The dependencies of the excess percent increase of Josephson currents of backbones arrays with respect to their reference as a function of the applied external magnetic field, namely $\Delta I(B)/I_{REF}(B) = [I_{BB}(B) - I_{REF}(B)]/I_{REF}(B) = I_{BB}/I_{REF} - 1$ are shown in Fig. 3. Here $I_{BB}(B)$ and $I_{REF}(B)$ are the values corresponding to currents for each specific field of backbone (subscript *BB*) array and reference array (subscript *REF*). These values are either obtained by reading their value at a given voltage or averaging over the respective vaules of current over the whole voltage span. The different methods, in general, provide consistent results. The error bars are within the squares visible in the figure. In this Figure (a) and (b) refer to different samples. We see that when the Josephson energy decreases (due to the decrease of the supercurrent induced by the magnetic field) all over the array there is gradual increase of $\Delta I/I_{REF}$ which displays almost a singularity for *B=27.4 G* when the graph-embedded arrays have current increases up to two times higher than the currents of the geometrically equivalent arrays. For further increases of the field the current of the reference array is essentially zero and a comparison does not make sense. In any case, the value of *27.4 G* is safely away from the value depressing completely the Josephson current, namely, the first zero of the diffraction pattern which is attained for *B=35G*.

The lines through the experimental data in Fig. 3 correspond to fits obtained from the functional dependency

$$\frac{\Delta I(B)}{I_{REF}(B)} = \frac{const}{\sqrt{B_c - B}} \qquad (1)$$

where *const* is a dimensional constant and $B_c=27.4G$ is the field value for which the vertical asymptote in the curve occurs. The quality of the fits, appreciable even by eye, is quantified by the fitting softwares returning the Coefficient Of Determination (*COD*) or *R-square* : the values of this coefficient are respectively 0.9998 and 0.9980 for the fits of Fig. 3a and Fig. 3b. These results indicate that we are likely in presence of a critical transition of the system. We note that the values of the currents of the two arrays in Fig. 3a were measured at different voltages *230 mV* and *200 mV* respectively for (a) and (b).

The results indeed were not much dependent on the specific voltage where the current was measured and results essentially identical can be obtained by averaging the currents all over the voltage span of the arrays. Indeed the relevance of the ratio $I_{BB}/I_{REF}$ came to our attention just when realizing that its value, for given magnetic field and temperature, is not dependent on the specific voltage value where it is evaluated. We note that in Fig. 3a, resp 3b, the currents of the reference arrays ($I_{REF}$) for the value of the field generating the noticeable increase (*27.4G*) are respectively *190 nA* and *205nA*. Those two values are not far from the *176 nA* that we estimated in the introduction for the current to which corresponds a Josephson energy equal to the thermal energy at *4.2K*.

In Figure 4a we show the current-voltage characteristics of the central finger array and its reference. Here we see that the current-voltage characteristic of the finger array (recall that we probe indeed two aligned fingers of the double comb) embedded in the graph structure and that of its geometrically equivalent, reference, array. The average current of the embedded array in this case is higher than that of the reference of about *500 nA* while the gap of each junction of the embedded array is higher of *45μV*, summing up to *4.5 mV* for the series connection of *100* junctions. In Fig. 4b we also show the magnetic field dependence of the normalized, and "excess" current of the embedded finger array and the line fitting the data is eq. 1, like for Fig. 3. In this case the value of excess current for each field was obtained combining the two set of data : the difference between the currents of the *IV* measured at *200 mV* and that obtained averaging the currents over all the voltage span. The two methods give very close results. The curve fitting the data is still obtained from (1) and returns an *R-square* of *0.9933*. In all the arrays we tested the increase/per junction of the gap in the finger arrays ranged in the *(40-80) mV* range. On the finger arrays we also put probes for testing individual junctions and those, tested independently, showed the same effects on Josephson currents and gap measured for the whole array [13].

We step now to the temperature characterization of the observed differences in the arrays. In Fig. 5a we show the temperature dependence of the excess current of the backbone, graph-embedded array, as a function of temperature $\Delta I(T)$ normalized to the value of the maximum Josephson current of the reference array at each given temperature $I_{REF}(T)$. Increasing the temperature the Josephson current (and energy) decreases and we observe now a pronounced increase at a temperature of *6.57 K* : for this value of the temperature the graph-embedded array current becomes about six times the current of the reference array, as shown in Fig. 5b. In this case the differences in current were measured at a voltage of *100 mV*, the point indicted by the arrows in the fugure. The line fitting the data in Fig. 5a corresponds to an equation similar to (1), namely

$$\frac{\Delta I(T)}{I_{REF}(T)} = \frac{const}{\sqrt{T_c - T}} \qquad (2)$$

Where *const* is a dimensional constant and $T_c$=*6.57 K* is the value of the vertical asymptote. The value and the dimension of the constant are naturally different from those of eq. 1. We see from the fitting that the inverse square root dependence on the independent variable provides, even in this case, is excellent and the "singular" increase of $\Delta I/I_{REF}$ has a "singularity" for a temperature of *6.57 K*. A *COD= 0.9995* was returned by MATLAB for this fit. For a temperature of *6.57 K* the maximum current of the reference array, measured at *100 mV*, is *270 nA*. For this value the zero bias Josephson energy results *8.9x10$^{-23}$ J* while the thermal energy, for *T=6.57 K*, is *9x10$^{-23}$J*. Thus, when the singular increase of the excess current occurs, the zero-bias Josephson energy in all the junctions equals the thermal energy. The evidence of Fig. 5 also leads us to conclude that the effect generating the increas of $\Delta I/I_{REF}$, like in Fig. 3 and Fig. 4b, is just the lowering of the Josephson energy and not other effects generated by field penetration in the junctions. The value of *6.57 K* is not far from the "singular" temperature that could be extracted in ref. 10 (see Fig. 3 of that paper), for samples with higher current densities, which was slightly above *7K*.

As we have just seen, for the temperature dependence the "singular" increase of the backbone embedded currents does occur when the thermal energy equals, within *1%*, the Josephson energy. However we have earlier seen, for the magnetic field dependence, that the singular increase occurs when the difference between Josephson and thermal energies is of the order of *7%* (for Fig. 3a) and *15%* (for Fig. 3b). We attribute the difference between the two cases, field and temperature dependence, to the fact that the uniformity of the temperature all over the junctions of the arrays is superior to the one achievable in terms of field uniformity.

We note that both $I_{BB}$ and $I_{REF}$ have a monotonous (decreasing) dependence both on *B* (all our measurements are relative to the first lobe of the diffraction pattern) and *T*. However, there is no obvious reason justifying the fact that this dependence does not preserve the value of the ratio between the two currents, increasing field or temperature. The fact that the ratio increases when the Josephson energies become close to the thermal energies (increasing field or temperature) indicates that something is happening to charge carriers in the embedded arrays due to this condition.

In Fig. 6a,b we show the gap differences between the graph-embedded array and the reference one for temperatures close to that generating the transition to the normal state. We see that, up to the transition temperature, the difference in gap remains well identifiable and we conclude that, according to the relation between gap and transition temperature in the BCS theory [14], there will be a slight difference even in transition temperature, the one of the embedded arrays being higher [13]. We could superimpose to the *IV* curves measured in zero field of Fig. 6 those measured with a high magnetic field and the traces would be literally superimposed and barely distinguishable. This means that magnetic fields of the order of tens of gauss have no effect on the gap differences. We also see in the figure that both subgap resistances (below the gap) and normal state resistance (above the gap) of the two arrays are absolutely identical meaning that the gap increase is an effect concerning the superconducting ground state of the arrays. In all the samples we tested the gap enhancement for the embedded arrays was ranging between *30 μV and 80 μV*.

While the theoretical model presented in refs. [1, 2, 3] was strictly related to the existence of bosons on the superconductive islands, the data herein reported indicate that the topological structures can condition even fundamental superconductive parametrs such as gap and condensation temperature [14] meaning that more physics can be extracted from graph arrays. It is known that Josephson current's amplitues can be linked to gap parameter through a BCS-originated equation [14, 15] and, perhaps, a relation might exist between the Josephson and gap "anomalies" we have herein identified. However, the relation might not be straightforward since the current anomalies that we have recorded over the years are somewhat stunning; in the above mentioned measurements reported in ref. 10 (Fig. 3), for example, the observed Josephson current increases were well above any limit that could be set by the Ambegaokar-Baratoff [15] equation.

## 3) CONCLUSIONS

The results herein reported confirm the reality of topology-induced "anomalies" in planar reticles of superconductive islands linked through Josephson tunnel junctions. We must admit that the phenomena visible in these graph-arrays are so way out of what one could expect based on usual superconductive, and Josephson, phenomenology, that we might still be putting aside interesting effects judging those artifacts or else. This has been the case of the gap increase of the embedded graphs observed ever since the first measurements performed [8] on comb-shaped graph reticles. Other stunning results have also been reported like, for example, those in Fig. 3 of ref. 10: in Fig. 3a of that paper, one can see that excess currents gets as high as allowed by the superconducting state and this phenomenon still deserves more careful analyses and measurements. In that particular sample the arrays had a higher Josephson critical currents and it would be interesting to further investigate the matter reported in the present paper for higher current densities.

Our data open interesting perspectives since the fact that properties of an array of superconducting islands are modified by specific topological connections is a result calling attention even on systems based on "traditional" superconductors. In high temperature superconductivity it is known that dimensionality and topology can play relevant roles, but, to our knowlege, specific experiments reporting on variations of gap parameter for structures like those we have characterized have never appeared in literature. However, a substantial amount of theoretical and experimental work has been dedicated over the past decades to arrays of Josephson junctions [16,17,18]. These papers all point toward effects induced by a collective behavior in arrays of Josephson junctions. Although evidences exist that a single Josephson junction might behave according to its equivalent electromagnetic circuit model down to tens of millikelvin tempertures [19], the data herein presented demonstrate that systems in which Josephson junctions are involved can display phenomena much characteristic of quantum statistics and hardly understandable in straight electromagnetic terms. It is possible that this existing background on theoretical and experimental investigation of statistical properties Josephson systems, and the present interest for topology-induced effects in condensed matter [20], might stimulate work and future developments.

This work was supported by the *FEEL* project of the Istituto Nazionale di Fisica Nucleare (Italy).

# REFERENCES

1) R. Burioni, D. Cassi, I. Meccoli, M. Rasetti, S. Regina, P. Sodano, and A. Vezzani, Europhysics Letters **52**, 251 (2000).

2) P. Buonsante, R. Burioni, D. Cassi, V. Penna, and A. Vezzani, Phys. Rev. **B70**, 224510 (2004).

3) I. Brunelli, G. Giusiano, F. P. Mancini, P. Sodano, and A. Trombettoni, J. Phys. B: At. Mol. Opt. **37**, S275-S286 (2004).

4) F. Fidaleo, J. Stat. Phys. **160**, 715 (2015).

5) T. Matsui, Infinite Dimensional Analysis, Quantum Probability and Related Topics **9**, 1 (2006).

6) R. Adami, E. Serra, and P. Tilli, Communications in Mathematical Physics **352**, 387 (2017).

7) M. L. Lyra, A. B. F. De Moura, I. N. de Oliveira, M. Serva, Phys. Rev. **E89**, 052133 (2014).

8) M. Cirillo, V. Merlo, R. Russo, M. G. Castellano, C. Cosmelli, A. Trombettoni, G. Giusiano, F. P. Mancini, and P. Sodano, "Spatial Bose-Einstein Condensation in Josephson Junctions Arrays" in *Quantum Computation in Solid State Systems*, B. Ruggiero, P. Delsing, C. Granata, Y. Pashkin, and P. Silvestrini eds., 147-153, Springer NY 2006.

9) P. Silvestrini, R. Russo, V. Corato, S. Rombetto, M. Russo, M. Cirillo, A. Trombettoni, and P. Sodano, Physics Letters **A370**, 499 (2007).

10) I. Ottaviani, M. Lucci, R. Menditto, V. Merlo, M. Salvato, M. Cirillo. F. Müller, T. Weimann, M. G. Castellano, F. Chiarello, G. Torrioli, and R. Russo, J. Phys.: Condens. Matter **26**, 215701 (2014); note that in this paper there is a typo in the expression of the height of the washboard potential ($I/I_c$ under the square root must be squared).

11) P. W. Anderson, Special Effects in Superconductivity, in *Lectures on the Many Body Problem*, Edited by E. R. Caianiello (Academic Press, New York, 1964), Vol. 2, pp. 113-135.

12) https://seeqc.com/wp-content/uploads/2019/12/SeeQCSDesignRules_S1.pdf

13) M. Lucci et al, *Dependence on topology of gap in superconductive networks*, to be published.

14) M. Tinkham, *Introduction to Superconductivity*, Dover (NY, 1996); T. Van Duzer and C. W. Turner *Principles of Superconducting Devices and Circuits*, Prentice-Hall (NJ, 1999).

15) V. Ambegaokar and A. Baratoff, Phys. Rev. Lett. **10**, 486 (1963).

16) G. Parisi, Journal of Mathematical Physics **37**, 5158 (1996).

17) J. Dziarmaga, A. Smerzi, W. H. Zurek, and A. R. Bishop, Phys. Rev. Lett. **88**, 167001 (2002).

18) S. P. Benz, M. S. Rzchowski, M. Tinkham, and C. J. Lobb, Phys. Rev. Lett. **64**, 693 (1990).

19) J. A. Blackburn, M. Cirillo, and N. Grønbech-Jensen, Physics Reports, **611**, 1-33 (2016).

20) Jing Wang and Shou-Cheng Zhang, Nature Materials **16**, 1062 (2019).

**FIGURE CAPTIONS**

1) (a) Sketch of a double comb array showing the biasing conditions for the current-voltage measurements. The lines connecting the squares (the superconducting islands) represent the connections through Josephson junctions. We test the distribution of carriers/current along the spine, or backbone directions and along the two aligned central fingers; (b) Final product of the fabrication procedure showing a portion of backbone array; the islands generating the backbone (crosses and big squares) have different geometrical shape but the same volume of carriers available for tunnelling in each junction. The arrow indicates the direction of the externally applied magnetic field.

2) (a) Current-voltage charcteristics of a *100* junctions biased backbone (black) array compared with its "reference array (red). We can clearly see that both Josephson currents and gap-sum voltage of the graph-embedded backbone array are larger;(b) Enlargement of the part of Fig. 2a close to the zero voltage axis showing the successive advancement of the gaps of the graph-embedded backbone array. Here we show both positive and negative parts of the characteristics demonstrating that the observed effects are not generated by one directional voltage offsets or else.

3) (a), (b) Magnetic field dependencies of the normalized current excess of the backbone arrays of double comb graph structures for two different samples. We see that the magnetic field, gradually reducing the Josephson currents of the arrays (and the relative coupling energy) provokes an enhancement of relative excess current. The statistical Coefficient Of Determination (*COD*) for the curve fittings is respectively *0.9998* in (a) and *0.998* in (b).

4) (a) Comparison between a graph-embedded double finger array and its reference; (b) magnetic field dependence of the excess current of the finger array shown in (a) as a function of the external magnetic field. COD for the curve fitting is *0.9933*.

5) (a) Dependence of excess current as a function of temperature. The fit is the a functional dependence which is analogous to that shown in Fig. 3 for the magnetic field behavior. As before, the line through the data is a fit following an inverse square root dependence; (b) The *IV* curves showing the noticeable difference between the currents of the backbone, graph-embedded, array (black) and its geometrical equivalent (red) at a temperature of *6.57 K*. The arrow indicate the value of voltage (*100 mV*) where currents are measured and we can see the noticeable difference between the current of the embedded back bone ($I_{BB}$) and its geometrically equivalent ($I_{REF}$). COD for the curve fit in (a) is *0.9995*.

6) Comparison of the gap-sum of the graph-embedded and reference arrays for temperatures close to the transition to the normal state: (a) *T=8 K* and (b) *T=8.5K*. The curves with higher gap values (indicated by BB) are relative to the graph-embedded arrays. Superimposing in (a) and (b) the curves obtained for a high magnetic field the latter would be hardly distinguishable from those we plot.

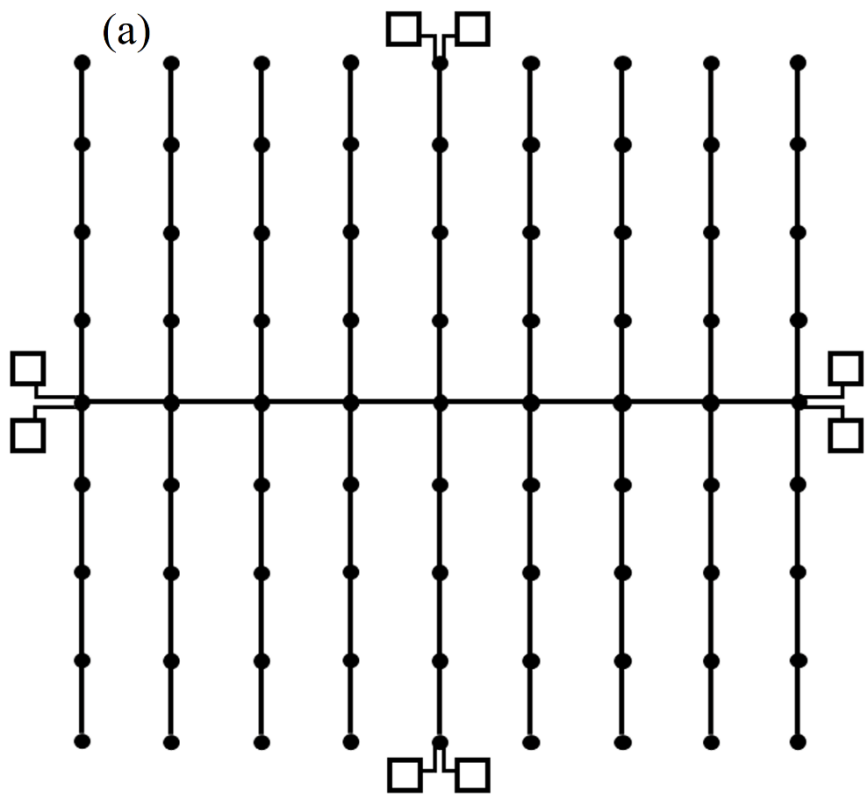

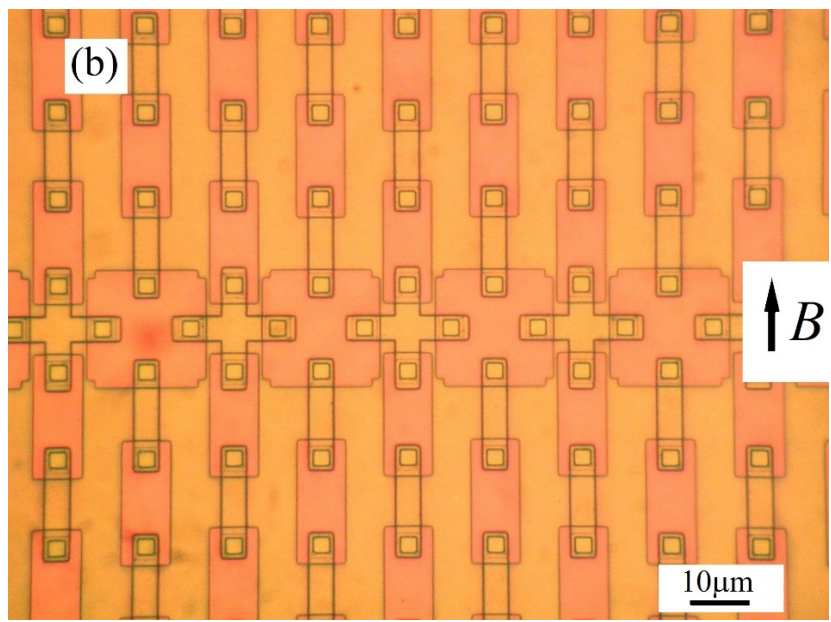

Fig. 1, M. Lucci et al.

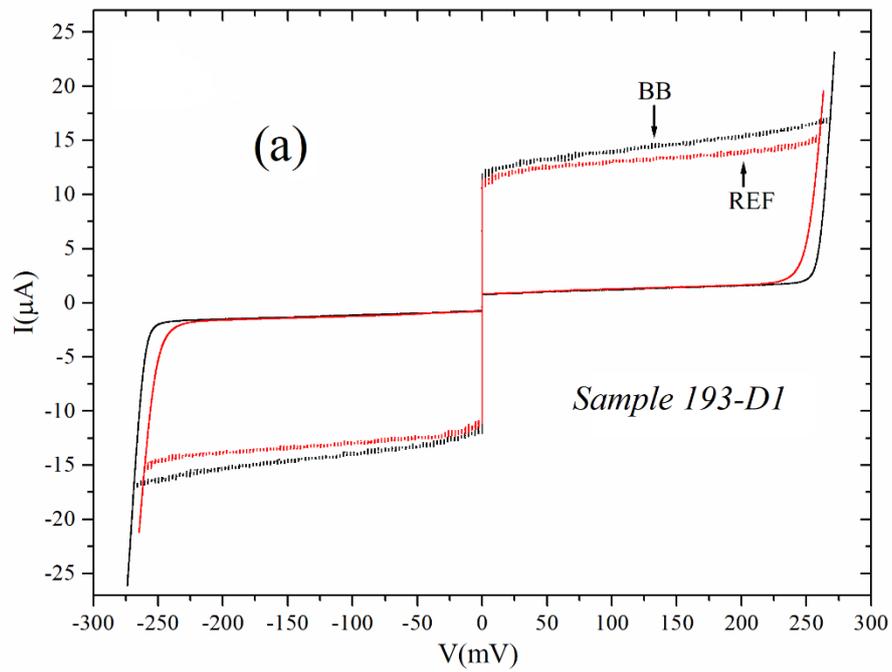

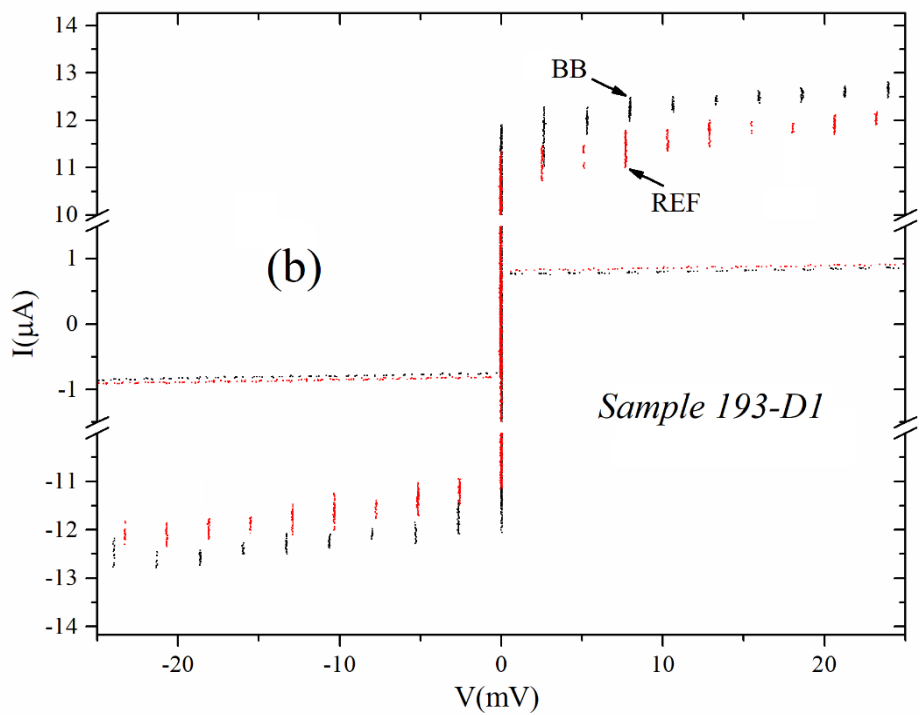

Fig. 2, M. Lucci et al.

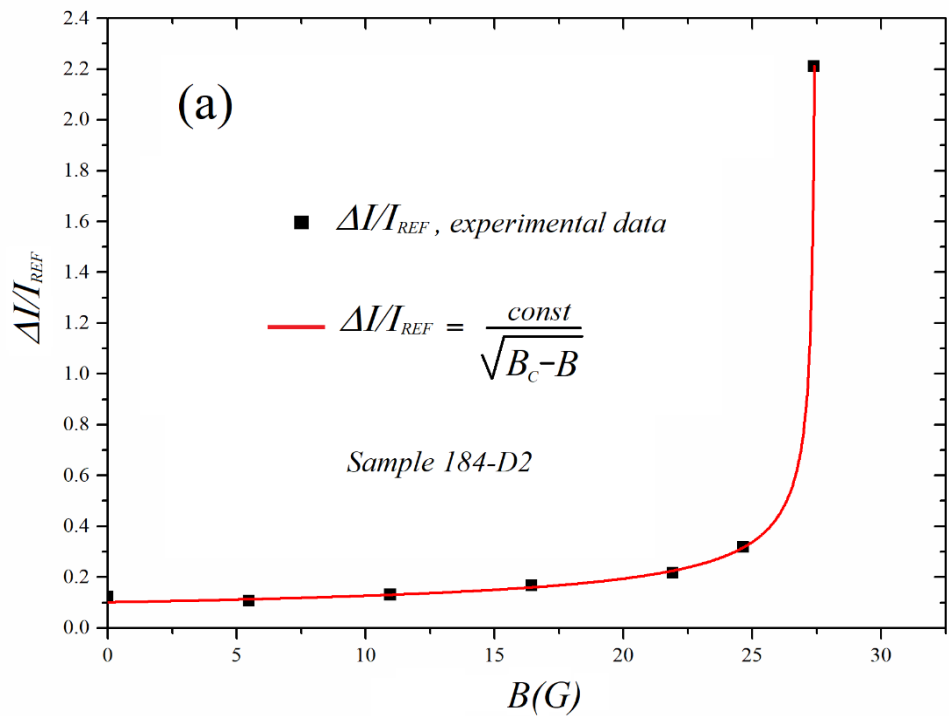

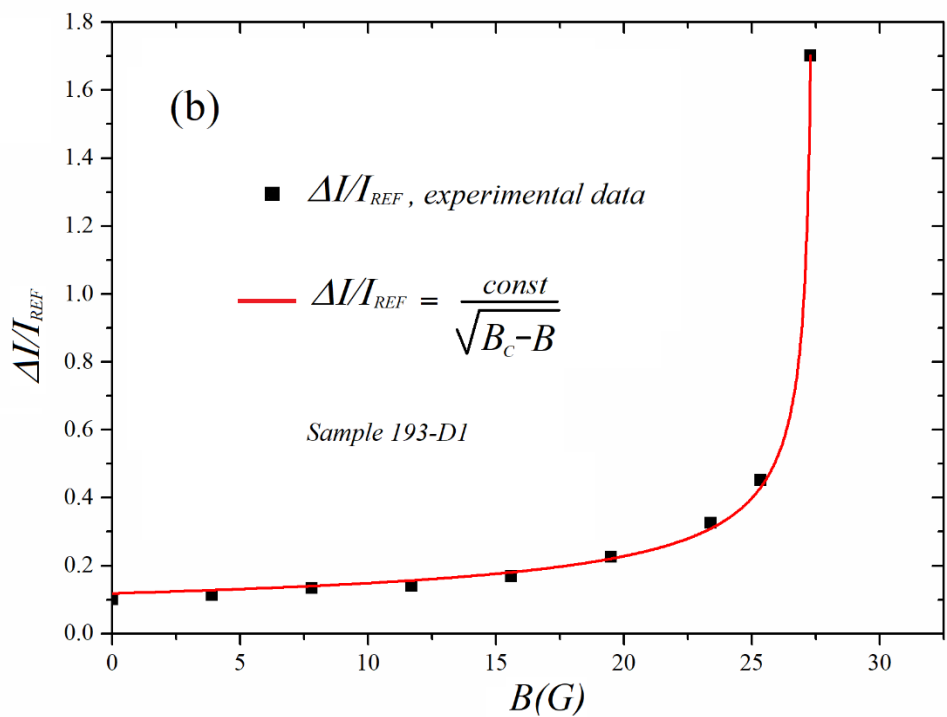

Fig. 3, M. Lucci et al.

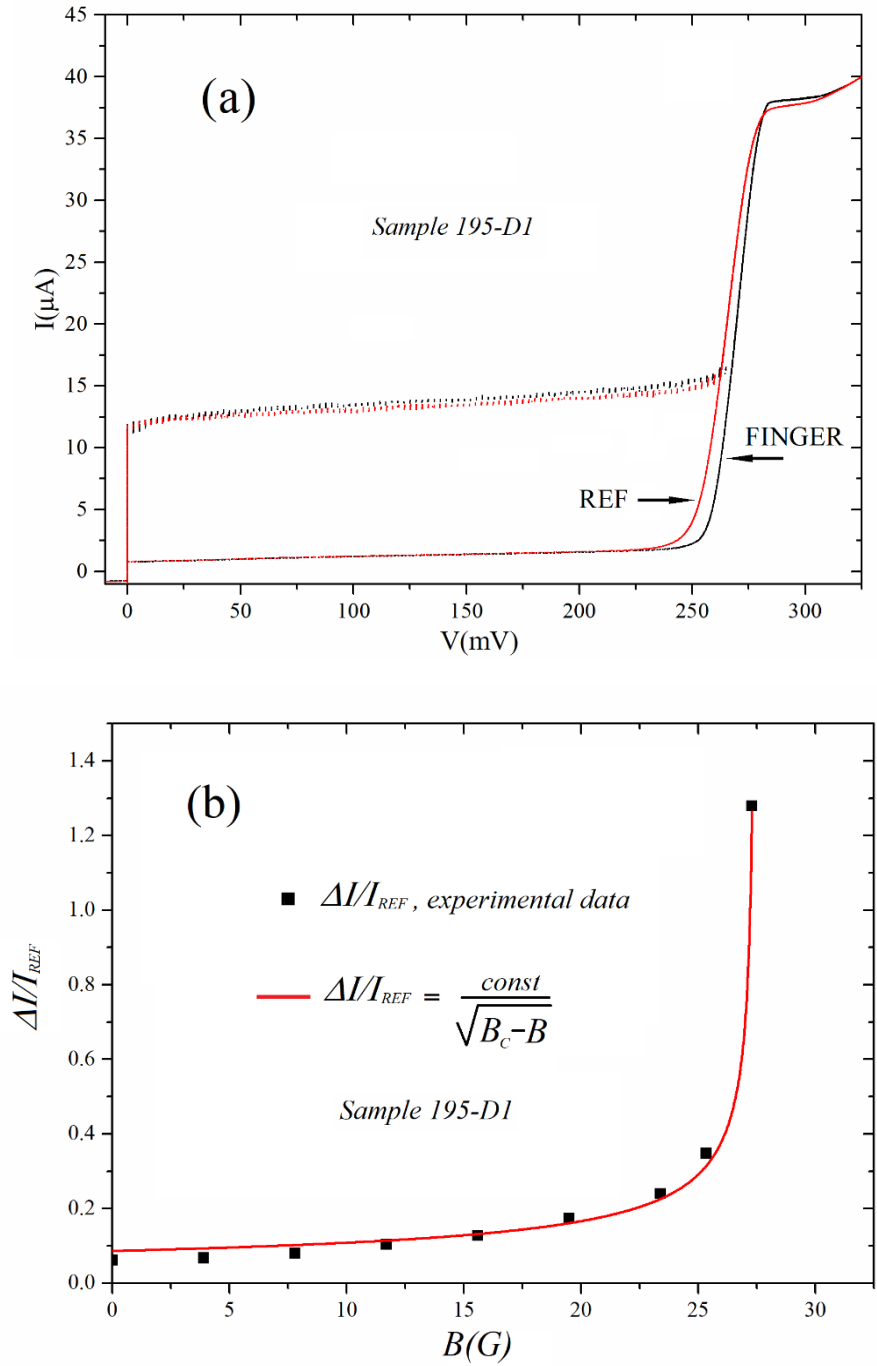

Fig. 4, M. Lucci et al.

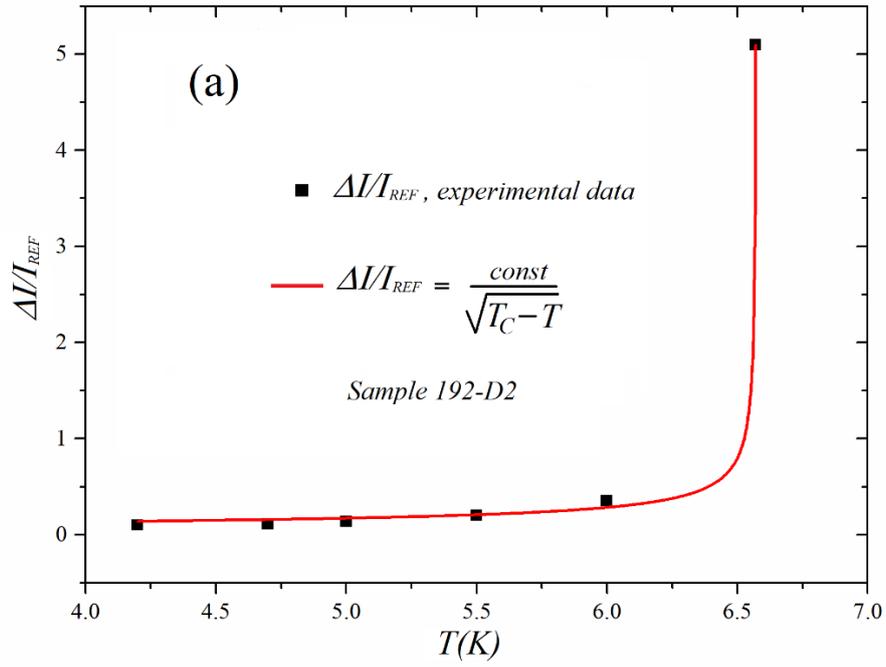

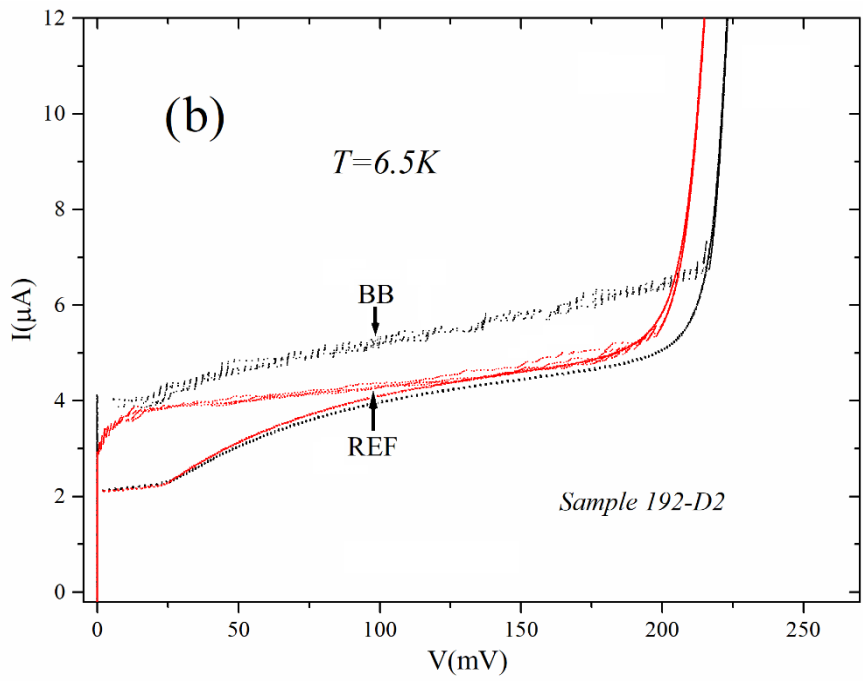

Fig. 5, M. Lucci et al.

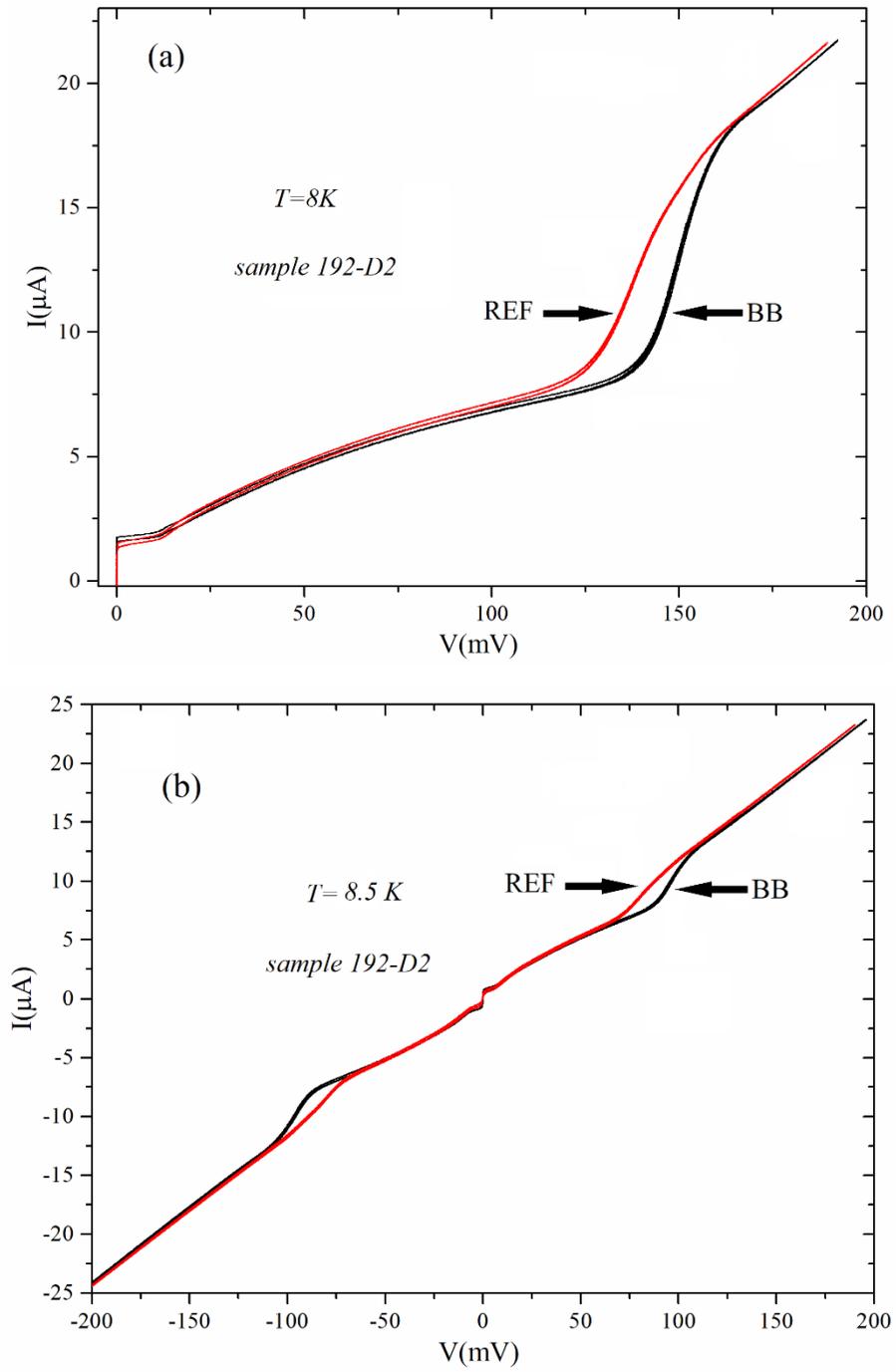

Fig. 6, M. Lucci et al.